\documentclass[preprint,showpacs,preprintnumbers,amsmath,amssymb]{revtex4}
\usepackage{graphicx}
\usepackage{dcolumn}
\usepackage{bm}
\begin{document}
\title {Lifetime of a quasiparticle in an electron liquid}
\author {Zhixin Qian}
\affiliation{Department of Physics  and State
Key Laboratory for Mesoscopic Physics, Peking University, Beijing 100871, China \\
and Department of Physics and Astronomy,
University of Missouri, Columbia, Missouri 65211, USA}
\author{Giovanni Vignale}
\affiliation{ Department of Physics and Astronomy, 
University of Missouri, Columbia, Missouri 65211, USA} 
\date{\today}
\begin{abstract}
We calculate the inelastic lifetime of an electron quasiparticle
due to Coulomb interactions in
an electron liquid at low (or zero) temperature in 
two and three spatial dimensions. The contribution of  ``exchange"  
processes is  calculated analytically and is shown to 
be non-negligible even in the high-density limit in two dimensions.  
Exchange effects must therefore be taken into account in 
a quantitative comparison between theory and experiment. 
The derivation in the two-dimensional case is presented in detail 
in order to clarify the origin of the disagreements that exist 
among the results of previous calculations, even the ones that only 
took into account  ``direct" processes.
\end{abstract}
\pacs{71.10.Ay, 71.10.-w, 72.10.-d}
\maketitle

\section{Introduction}
The calculation of the inelastic scattering lifetime of an excited  
quasiparticle in an electron liquid, due to Coulomb
interactions, is a fundamental problem in quantum many-body theory.  
According to the Landau theory of Fermi liquids~\cite{pines} 
the inverse lifetime  of an electron quasiparticle of  
energy $\xi_p$ (relative to the Fermi energy $E_F$) at 
temperature $T$ in a three-dimensional (3D)  electron liquid should scale as  
\begin{equation}\label{Landauscaling3D}
\frac{\hbar}{\tau_e} \propto \left\{ \begin{array}{l}
\left(\frac{\xi_p}{E_F}\right)^2~~~~~~k_BT \ll \xi_p \ll E_F\\
\left(\frac{k_BT}{E_F}\right)^2~~~~\xi_p \ll k_BT \ll E_F~,
\end{array}~~~~~~~~~~~~~(3D)\right.
\end{equation} 
where $k_B$ is the Boltzmann constant.    
In a two-dimensional (2D) electron liquid the above 
dependencies are modified as follows~\cite{giuliani}:
\begin{equation}\label{Landauscaling2D}
\frac{\hbar}{\tau_e} \propto \left\{ \begin{array}{l}
\left(\frac{\xi_p}{E_F}\right)^2
\ln\frac{E_F}{\xi_p}~~~~~~~k_BT \ll \xi_p \ll E_F\\
\left(\frac{k_BT}{E_F}\right)^2
\ln\frac{E_F}{k_BT}~~~~\xi_p \ll k_BT \ll E_F~.
\end{array}~~~~~~(2D)\right.
\end{equation}

Besides its obvious importance for the foundations of the 
Landau theory of Fermi liquids~\cite{pines},  
the inelastic lifetime also plays a key role in our understanding 
of certain transport phenomena, such as weak localization in disordered metals.  
In this case,  the distance an electron diffuses during 
its inelastic lifetime provides the natural upper cutoff for 
the scaling of the conductance, and thus determines 
the low-temperature behavior of the latter~\cite{choi,yacoby,kawaji}. 

During the past decade some newly developed experimental techniques, 
combined with the ability to produce high-purity 2D electron liquids 
in semiconductor quantum wells have enabled  experimentalists to 
attempt for the first time a direct determination of the intrinsic 
quasiparticle lifetime, i.e., the lifetime that arises purely from 
Coulomb interactions in a low-temperature, clean electron 
liquid~\cite{berk,murphy,slutzky}.  In Refs.~\cite{murphy,slutzky}, 
for example,  the quasiparticle lifetime was  extracted directly from the 
width of the electronic spectral function obtained  from a measurement of 
the tunneling conductance between two quantum wells. 
In the case of large wells separation, like the ones ($175 \AA \sim 340 \AA$) 
studied in Ref.~\cite{slutzky}, the couplings between electrons 
in different well are weak and can be ignored.   For such weakly-coupled wells, 
the lifetime  is principally due to interactions among electrons in 2D, 
while the contribution of the impurities is relatively small. 

In spite of these wonderful advances, a quantitative comparison 
between theory and experiment remains very difficult.  There are several 
reasons for this to be so.   First of all, the 2D samples studied in 
the experiments are not yet sufficiently ``ideal", namely disorder 
and finite width effects still play a non-negligible role: as a result, 
the measured lifetimes are typically found to be considerably shorter than 
the theoretically calculated ones.   Secondly, the electronic density in 
these systems falls in a range in which the traditional high-density/weak-coupling  
approximations~\cite{pines,quinn,ritchie,kleinman,penn}, are not  
really justified.   Finally,  there is  still confusing disagreement 
among various theoretical results 
in 2D \cite{chaplik,hodges,giuliani,fukuyama,jungwirth,fasol,zheng,menashe,reizer}, 
even in the random phase approximation (RPA).

This paper is devoted to a critical analysis of the last question, i.e., 
specifically, we calculate {\it analytically} the constants of 
proportionality in the relations~(\ref{Landauscaling3D}) 
and~(\ref{Landauscaling2D}) in the weak coupling regime, 
and try to clear up the differences that exist among the 
results of different published calculations. One particular aspect of 
the confusion is the widespread belief that the Fermi golden rule calculation 
of the lifetime, based on the RPA screened interaction,  
is exact in the high density/weak coupling limit.  
In fact, this is only true in 3D, but not in 2D.  
To our knowledge, this fact was first recognized by Reizer and Wilkins~\cite{reizer}, 
who introduced what they called  ``non golden rule processes", 
i.e., exchange processes in which the quasiparticle is replaced in the final state 
by one of the particles of the liquid.  In point of truth, these processes 
are still described by the Fermi golden rule, provided one recognizes 
that the initial and final states are  Slater determinants, 
rather than single plane wave states.  In three dimensions, such exchange 
contributions to the lifetime were calculated (numerically) 
in Refs.~\cite{kleinman,penn},  but they are easily shown to become 
irrelevant in the high-density limit.  In 2D, by contrast, 
the exchange contribution remains of the same order as the direct contribution 
even in the high density limit.  Reizer and Wilkins found the exchange 
contribution to reduce to $\frac{1}{2}$ of the direct one 
(with the opposite sign) in the high-density limit, while we find  it 
here to be only $\frac{1}{4}$ of the direct contribution in the same limit.  
More generally, we give an analytical evaluation of both the ``direct" and  
the ``exchange" contributions vs density, for boh  $k_BT \ll  \xi_p$ and $\xi_p\ll k_BT$.

The rest of this paper is organized as follows.  
In Sect. II, we provide the general formulas for  $\frac{\hbar}{\tau_e}$ 
including exchange processes.  We then devote Sect. III
to the analytical calculation of  $\frac{\hbar}{\tau_e}$ in 3D; 
and Sect. IV to the same calculation in 2D.  The 2D calculation is 
presented in greater detail in order to explain the origin of the 
disagreements among the results of previous calculations.   
We explain the reason for the much stronger impact of exchange on 
the lifetime in 2D than in 3D at high density.  
Section V presents a comparison between the present theory and 
the experimental data of Ref.~(\cite{murphy}) and summarizes the ``state of the art".

\section{General formulas}

We consider an excited quasiparticle with momentum ${\bf p}$ and
spin $\sigma$. Its inverse inelastic lifetime due to 
the electron-electron interaction
is a sum of two terms, corresponding to the contributions from
the ``direct" and ``exchange" processes, respectively,
\begin{eqnarray} \label{sum}
\frac{1}{\tau_e(\xi_p, T)}  = \frac{1}{\tau_{ \sigma}^{(D)}}
+ \frac{1}{\tau_{\sigma}^{(EX)}} ,
\end{eqnarray}
where $\xi_p \equiv \frac{p^2}{2 m}- \mu$ is the free-particle energy 
measured from the chemical potential $\mu$. We use `D' to denote 
the ``direct" term,  and `EX' the ``exchange" term.

\begin{figure}\label{fig1}
\includegraphics[width=12cm]{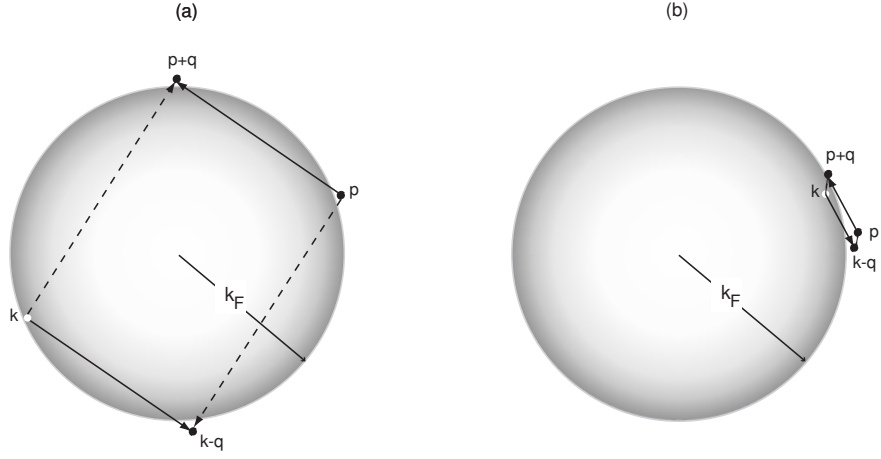}
\caption {(a) A typical  scattering process between electrons 
of the same spin orientation near the Fermi surface 
has contributions from both a ``direct"  (solid line) and 
an ``exchange" (dotted line)  term;  
(b) A special class of low momentum transfer processes 
gives the leading-order contribution to the 
scattering amplitude in 2D at high density.}
\vspace{1pt}
\end{figure}

Making use of the Fermi golden rule, we get
\cite{pines},
\begin{eqnarray} \label{tauD00}
\frac{1}{\tau_{  \sigma}^{(D)}}
= 2 \pi &&\sum_{{\bf k}, {\bf q}} \sum_{\sigma'}
W^2({\bf q})
{\bar n}_{{\bf p} + {\bf q}\sigma} n_{\bf k \sigma'}
{\bar n}_{{\bf k} - {\bf q} \sigma'}   \nonumber \\
&& \times
\delta (\xi_p + \xi_{{\bf k} \sigma'}
- \xi_{{\bf k} - {\bf q} \sigma'} -
\xi_{{\bf p}  + {\bf q} \sigma}) ,
\end{eqnarray}
and
\begin{eqnarray} \label{tauex00}
\frac{1}{\tau_{  \sigma}^{(EX)}}
=&& -2 \pi \sum_{{\bf k}, {\bf q}}
W({\bf p}-{\bf k}+ {\bf q}) W({\bf q})
{\bar n}_{{\bf p}+ {\bf q} \sigma}{\bar n}_{{\bf k} - {\bf q}\sigma}
n_{{\bf k} \sigma} \nonumber \\
&& \times
\delta (\xi_p + \xi_{{\bf k}  \sigma} - \xi_{{\bf k}-{\bf q}\sigma}  -
\xi_{{\bf p} +{\bf q} \sigma}) ,
\end{eqnarray}
where $W({\bf q})$ is the effective interaction between two quasiparticles,
$n_{{\bf k} \sigma} = \frac{1}{e^{\beta\xi_{\bf k}}+1}$  the 
Fermi-Dirac distribution function at temperature $\beta=\frac{1}{k_BT}$,  and
we have set $\hbar=1$. 
The $\delta$-functions ensure the conservation of the energy in the
collisions. Obviously, from Eqs. (\ref{tauD00}) and (\ref{tauex00}), one can
see that the contribution from the ``exchange" process tends to cancel that from
the ``direct" process.

As can be seen from Eq. (\ref{tauD00}), there are two types of
collisions contributing to the ``direct" term, 
the collisions with same-spin
electrons ($\sigma' =\sigma$), and those with opposite-spin electrons
($\sigma' =-\sigma$). We denote the former $1/\tau_{\sigma \sigma}$, and
the latter $1/\tau_{\sigma {\bar \sigma}}$, where ${\bar \sigma}=-\sigma$.
It can be easily shown that
\begin{eqnarray}
\frac{1}{\tau_{ \sigma \sigma}^{(D)}} \geq - \frac{1}{\tau_{  \sigma}^{(EX)}} .
\end{eqnarray}
In the paramagnetic state, one evidently has
\begin{eqnarray}  \label{equation}
\frac{1}{\tau_{\sigma \sigma}^{(D)}} 
=\frac{1}{\tau_{\sigma \bar \sigma}^{(D)}} .
\end{eqnarray}
Therefore,
\begin{eqnarray}  \label{exact1}
\frac{1}{2 \tau_{\sigma }^{(D)}} \geq 
- \frac{1}{\tau_{\sigma}^{(EX)}} . 
\end{eqnarray}

The effective interaction $W({\bf q})$ between quasiparticles 
is short-ranged compared to the bare Coulomb potential due to
the screening effects from the remaining electrons. Such screening
effects are normally characterized by a screening
wave vector $k_s$.  Following this practice  we  approximate
\begin{equation}
W({\bf q}) = \left\{ \begin{array}{l} \frac{4 \pi e^2}{q^2 +k_s^2}~~~~~(3D)~\\
\frac{2 \pi e^2}{q +k_s}~~~~~~(2D)~, 
\end{array}\right.
\end{equation}
where 
\begin{equation}
k_s = \left\{ \begin{array}{l} \sqrt{\frac{4k_F}{\pi a_0}}~~~~~(3D)\\
~~~\frac{2}{a_0}~~~~~~~(2D)~,\end{array}\right.
\end{equation}
and $k_F$ and $a_0$ are the Fermi wave vector 
and the Bohr radius, respectively.
At very low density, the screening wave vector becomes
much larger than the Fermi wave vector. It can be shown that, in this limit,
\begin{eqnarray} 
\frac{1}{\tau_{ \sigma \sigma}^{(D)}} 
= - \frac{1}{\tau_{ \sigma}^{(EX)}} ,
\end{eqnarray}
or, in other words, by using Eq. (\ref{equation}), 
\begin{eqnarray}  \label{exact2}
\frac{1}{2 \tau_{ \sigma }^{(D)}} =
- \frac{1}{\tau_{ \sigma}^{(EX)}} .
\end{eqnarray}
Eq. (\ref{exact1}) and the low density limit result of 
Eq. (\ref{exact2}) are exact results,
which, to the best of our knowledge, were not explicitly established
before.

In what follows we will only consider the  case of the paramagnetic  
electron liquid, which allows us to trivially dispose of the spin indices.  
Furthermore, by making the change of  variable  ${\bf k} \to {\bf k}+{\bf q}$ in 
the momentum summation in Eq. (\ref{tauex00}), 
and correspondingly, ${\bf k} \to -{\bf k}$
in Eq. (\ref{tauD00}), we rewrite Eqs. (\ref{tauD00}) and (\ref{tauex00}) as
\begin{eqnarray} \label{tauD}
\frac{1}{\tau^{(D)}}
= 2 \pi &&\sum_{{\bf k}, {\bf q}} \sum_{\sigma'}
W^2({\bf q})
{\bar n}_{{\bf p} + {\bf q}} n_{\bf k }
{\bar n}_{{\bf k} + {\bf q} }   \nonumber \\
&& \times
\delta (\xi_p + \xi_{\bf k }
- \xi_{{\bf k} + {\bf q} } -
\xi_{{\bf p}  + {\bf q} }) ,
\end{eqnarray}
and
\begin{eqnarray} \label{tauex0}
\frac{1}{\tau^{(EX)}}
=&& -2 \pi \sum_{{\bf k}, {\bf q}}
 W({\bf p}-{\bf k}) W({\bf q})
{\bar n}_{{\bf p}+ {\bf q} }{\bar n}_{\bf k}
n_{{\bf k}+{\bf q} } \nonumber \\
&& \times
\delta (\xi_p + \xi_{{\bf k} +{\bf q} } - \xi_{\bf k}  -
\xi_{{\bf p} +{\bf q} }) .
\end{eqnarray}
By using the identity,
\begin{eqnarray}
\frac{\Im m \chi_0(q, \omega)}{1 - e^{-\beta \omega}} = - 2 \pi
\sum_{\bf k} n_{\bf k} {\bar n}_{{\bf k} + {\bf q}}
\delta (\omega + \xi_{\bf k} - \xi_{{\bf k} + {\bf q}}) ,
\end{eqnarray}
where $ \chi_0(q, \omega)$ is the  Lindhard function
(i.e., the density-density response function of the non interacting electron gas),
we rewrite $1/\tau^{(D)}$ in Eq. (\ref{tauD}) as
\begin{eqnarray} \label{tauD2}
\frac{1}{\tau^{(D)}}
&& =  -2 \int_{-\infty}^\infty d \omega \frac{1}
{[1 + e^{\beta (\omega - \xi_p)}][1 - e^{-\beta \omega}]} \nonumber \\ 
&& \times \sum_{\bf q}
W^2({\bf q}) 
\delta (\omega - \xi_p + \xi_{{\bf p} + {\bf q}})
Im \chi_0(q, \omega) .
\end{eqnarray}
In obtaining Eq. (\ref{tauD2}),
we have also used the fact that
\begin{eqnarray}
{\bar n}_{{\bf p} + {\bf q}} \delta (\omega - \xi_p + \xi_{{\bf p} + {\bf q}})
=\frac{1}{1 + e^{\beta (\omega - \xi_p)}} 
\delta (\omega - \xi_p + \xi_{{\bf p} + {\bf q}}) .  \nonumber \\
\end{eqnarray}
Similarly, one has
\begin{eqnarray}  \label{tauX2}
\frac{1}{\tau^{(EX)}}
&& =  -2 \pi \int_{- \infty}^\infty d \omega \frac{1} 
{1 + e^{\beta (\omega - \xi_p)}} 
\sum_{{\bf k}, {\bf q}} W({\bf q}) {\bar n}_{\bf k} n_{{\bf q}+{\bf k}} 
 \nonumber \\
&& \times \delta(\omega - \xi_{\bf k} + \xi_{{\bf q} +{\bf k}}) 
\delta (\omega - \xi_p + \xi_{{\bf q}+ {\bf p}})
W({\bf p} - {\bf k}) .   \nonumber \\
\end{eqnarray}

The fact that $1/\tau^{(D)}$ and 
$1/\tau^{(EX)}$ depend only of the magnitude
of ${\bf p}$ allows us to average
over the unit vector of ${\bf \hat p}  = \frac{{\bf p}}{p}$ on the right hand side of 
Eqs. (\ref{tauD2}) and (\ref{tauX2}). To this end, we define
\begin{eqnarray}
\Omega_{\pm} (q) \equiv \pm \frac{pq}{m} - \frac{q^2}{2m},
\end{eqnarray}
and use the fact that
\begin{eqnarray}  \label{identity}
\frac{1}{2^{d-1} \pi} \int d {\bf \hat p} 
&& \delta (\omega - \xi_p + \xi_{{\bf q}+ {\bf p}}) 
= \Theta(p, q)  \nonumber \\
&& \times 
\theta(\Omega_+ (q) - \omega) 
\theta(\omega - \Omega_- (q))~,
\end{eqnarray}
where $\theta(x) =1$ 
for $x>0$, and $\theta(x)=0$ for $x\leq 0$, and  
\begin{equation}
\Theta(p, q) = \left\{\begin{array}{l}\frac{m}{2 pq}~,~~~~~~~~~~~~~~~~~~~~~(3D)\\
\frac{2m}{\pi \sqrt{4 p^2 q^2 - (2 m \omega +q^2)^2}}~,~~~(2D)~.
\end{array}\right.
\end{equation}
Therefore $1/\tau^{(D)}$ can be rewritten as
\begin{eqnarray} \label{tauD3}
\frac{1}{\tau^{(D)}}
&=&-2 \int_{-\infty}^\infty d \omega \frac{1}
{[1 + e^{\beta (\omega - \xi_p)}][1 - e^{-\beta \omega}]}\nonumber \\ &\times& \sum_{\bf q}
W^2(q) \Im m \chi_0(q, \omega)  \Theta(p, q) \theta(\Omega_+ (q) - \omega)
\theta(\omega - \Omega_- (q))~.
\end{eqnarray}
We note that this equation is  not restricted to the
regime of $k_BT \ll E_F$, but holds for arbitrary temperature.

In this paper, we are only interested in the case
that $k_BT \ll E_F$, and therefore the Fermi
energy $E_F$ is always well defined and $E_F \simeq \mu$.
To perform the average over ${\bf \hat p}$ in  
Eq.~(\ref{tauX2}), we use the fact that, for
$k_B T, \xi_p \ll E_F$, the contribution to $1/\tau^{(EX)}$
only arises from the region in which $\xi_{\bf k},
|\omega| \ll E_F$.  Furthermore, the first $\delta$-function 
in Eq.~(\ref{tauX2}) fixes the angle between ${\bf k}$ and ${\bf q}$ to 
be such as to satisfy the 
condition $\xi_{\bf k}-\xi_{{\bf q}+ {\bf k}} = \omega \approx 0$.  
With this in mind, one obtains
\begin{eqnarray}\label{exaverage}
\frac{1}{2^{d-1} \pi} && \int d {\bf \hat p}
\delta (\omega - \xi_p + \xi_{{\bf q}+ {\bf p}}) 
W({\bf p} - {\bf k})      \nonumber \\
&& =\Phi(p, q) \theta(\Omega_+ (q) - \omega)
\theta(\omega - \Omega_- (q)) ,
\end{eqnarray}
where
\begin{equation}
\Phi(p, q) = \left\{\begin{array}{l}
\frac{m}{2 pq k_s} \frac{4 \pi e^2}
{\sqrt{k_s^2 + 4 k_F^2 - q^2}}~,~~~~~~~~~~~~~~~~~~~~~~~~~~~~(3D)\\
\frac{ m e^2 }
{\sqrt{p^2 q^2 - (m \omega + q^2/2)^2}}\left [\frac{1}{k_s} 
+ \frac{1}{\sqrt{4 k_F^2 -q^2} +k_s} 
\right] ~,~~(2D)~.
\end{array}\right.
\end{equation} 
A detailed derivation of this key result is presented in the Appendix.  Thus finally
\begin{eqnarray}  \label{tauX3}
\frac{1}{\tau^{(EX)}}
&=& \int_{-\infty}^\infty d \omega \frac{1}
{[1 + e^{\beta (\omega - \xi_p)}][1 - 
e^{-\beta \omega}]} \nonumber \\
&  \times& \sum_{\bf q}
W(q)  \Im m \chi_0(q, \omega) \Phi(p, q) \theta(\Omega_+ (q) - \omega)
\theta(\omega - \Omega_- (q)) .
\end{eqnarray}

\section{The inverse lifetime in 3D}
The theory of the electron inelastic lifetime 
in 3D is rather well established~\cite{quinn,ritchie} at zero temperature.
However, no analytical expression including 
the exchange has been presented so far, even
though Kleinman~\cite{kleinman}, and later Penn~\cite{penn}, 
have reported numerical calculations of the exchange contribution.  
This deficiency is remedied in the present section. Our calculation is
done at nonzero temperature,  with zero temperature as a special case.

In 3D, Eq. (\ref{tauD3}) becomes
\begin{eqnarray} \label{tauD3again}
\frac{1}{\tau^{(D)}}
&=&-\frac{m}{2 (2 \pi )^3 p} \int_{-\infty}^\infty
d \omega \frac{2}{[1+e^{\beta (\omega - \xi_p)}]
[1 - e^{- \beta \omega} ] } \nonumber  \\
&\times& \int d {\bf q}
W^2 (q) \frac{1}{q} \Im m \chi_0(q, \omega) \theta(\Omega_+(q) - \omega) 
\theta(\omega - \Omega_-(q)) .
\end{eqnarray}
We are interested in the case that $k_BT, \xi_p \ll E_F $.
Therefore we only need consider the region of $\omega \ll E_F $, in which,
\begin{eqnarray} \label{chi3D}
Im \chi_0(q, \omega)= -\frac{m^2 \omega}{2 \pi q} \theta(2k_F -q) .
\end{eqnarray}
Substituting Eq. (\ref{chi3D}) into (\ref{tauD3again}) leads to
\begin{eqnarray}
\frac{1}{\tau^{(D)}}
= \frac{m^3}{(2 \pi)^3 p} &&\int_{-\infty}^{\infty} d \omega
\frac{ 2 \omega} 
{[1 + e^{\beta (\omega - \xi_p)}]
[1 - e^{-\beta \omega}]} \nonumber \\
&& \times \int_0^{2 k_F} d q W^2(q) .
\end{eqnarray}
The integrations over $q$ and $\omega$ can be carried through,
and one obtains 
\begin{eqnarray}  \label{tau3Dfina}
\frac{1}{\tau^{(D)}}
= \frac{m^3 e^4}{\pi p k_s^3} \frac{\pi^2 k_B^2T^2 +\xi_p^2}
{1+e^{- \beta \xi_p}} \biggl [ \frac{\lambda}{\lambda^2 +1}
+ \tan^{-1} \lambda \biggr ] ,
\end{eqnarray}
where $\lambda = \frac{2 k_F}{k_s}$. 

Next we move to evaluate the contribution from the exchange process.
In 3D, Eq. (\ref{tauX3}) becomes
\begin{eqnarray}
\frac{1}{\tau^{(EX)}}
&=&\frac{\pi e^2 m}{(2 \pi)^3pk_s}
\int_{-\infty}^\infty d \omega 
\frac{2}{1 + e^{\beta (\omega - \xi_p)}]
[1 - e^{-\beta \omega}]}  \nonumber \\
&\times&  \int d {\bf q} W(q) \frac{\Im m \chi_0(q, \omega)}
{q \sqrt{k_s^2 + 4 k_F^2 - q^2}}\theta(\Omega_+(q) - \omega) \theta(\omega - \Omega_-(q)) .
\end{eqnarray}
By using Eq. (\ref{chi3D}), one has
\begin{eqnarray}
\frac{1}{\tau^{(EX)}}
=- &&\frac{m^3 e^2 }{(2 \pi)^2pk_s}
\int_{-\infty}^\infty d \omega 
\frac{2 \omega}{[1 + e^{\beta (\omega - \xi_p)}]
[1 - e^{-\beta \omega}]}  \nonumber \\
&& \times \int_0^{2k_F} d  q W(q) \frac{1}{ \sqrt{k_s^2 + 4 k_F^2 - q^2}} .
\end{eqnarray}
After carrying out the integrations, one obtains the final
result,
\begin{eqnarray}
\frac{1}{\tau^{(EX)}}
&&=-\frac{m^3 e^4}{ \pi p k_s^3} \frac{\pi^2 k_B^2T^2 + \xi_p^2}
{1 + e^{-\beta \xi_p} } \nonumber  \\
&& \times \frac{1}{\sqrt{\lambda^2 +2 }}
\biggl [ \frac{\pi}{2} - \tan^{-1}
\biggl ( \frac{1}{\lambda}
\sqrt{\frac{1}{\lambda^2 +2 }} \biggr ) \biggr ] .
\end{eqnarray}

We plot the ratio of $1/\tau^{(EX)}$ to $1/\tau^{(D)}$ vs the Wigner-Seitz 
radius $r_s$ in Fig.~2. 
Notice that at very high density, 
$|1/\tau^{(EX)}| \ll |1/\tau^{(D)}|$,
and the direct-process-only theory is then relatively good. On the other
hand, at low density, $1/\tau^{(EX)} = - 1/2 \tau^{(D)}$, which
agrees with the general conclusion of Eq.~(\ref{exact2}). 
The contribution from exchange processes 
therefore cannot be ignored in most density range.

\begin{figure}\label{fig3}
\includegraphics[width=7cm]{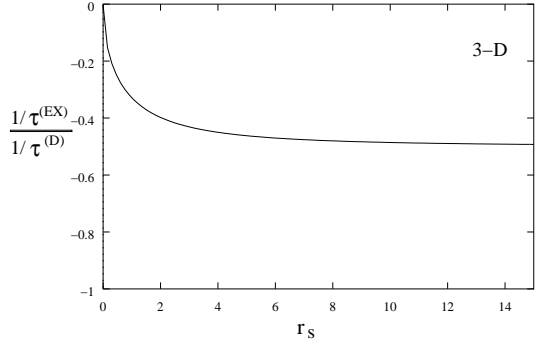}
\caption {The ratio of $1/\tau^{(EX)}$ 
over $1/\tau^{(D)}$ via $r_s$ in 3D.}
\vspace{1pt}
\end{figure}

In the limiting case of small
excitation energy, $\xi_p \ll k_BT \ll E_F$, Eq. (\ref{tau3Dfina})
reduces to
\begin{eqnarray}
\frac{1}{\tau^{(D)}}
= \frac{\pi m^3 e^4}{2 p k_s^3}  k_B^2T^2
\biggl [ \frac{\lambda}{\lambda^2 +1}
+ \tan^{-1} \lambda \biggr ] .
\end{eqnarray}

In the opposite  of very low temperature, 
$k_BT \ll \xi_p \ll E_F$, one has
\begin{eqnarray} \label{tau3DT=0}
\frac{1}{\tau^{(D)}}
= \frac{m^3 e^4}{\pi p k_s^3} \xi_p^2
\biggl [ \frac{\lambda}{\lambda^2 +1}
+ \tan^{-1} \lambda \biggr ] .
\end{eqnarray}
In the high density limit, $\lambda \to \infty$, Eq. (\ref{tau3DT=0})
becomes
\begin{eqnarray}
\frac{1}{\tau^{(D)}}
= \frac{m^3 e^4}{2 p k_s^3} \xi_p^2 ,
\end{eqnarray}
a result obtained earlier by Quinn and Ferrell~\cite{quinn}.

\section{The inverse lifetime in 2D}

As mentioned in the introduction, there is still some  
disagreement among the results of previous calculations of  $1/\tau_e$ in 2D. 
The main purpose of this section is to exactly evaluate the 
prefactors of $1/\tau^{(D)}$ and $1/\tau^{(EX)}$
in 2D, and at the same time attempt to clarify the origin of 
those disagreements. We present our derivations in the two different 
regimes of $k_BT \ll \xi_p \ll E_F$ and $\xi_p \ll k_BT \ll E_F$ separately.  
For greater clarity, we also show our derivations for the ``direct" and  
``exchange" contributions in separate subsections. 

\subsection{$k_BT \ll \xi_p$: Direct process}
In 2D, for $k_BT \ll \xi_p
\ll E_F $, Eq. (\ref{tauD3}) becomes
\begin{eqnarray}\label{threeregions}
\frac{1}{\tau^{(D)}}
=&& - \frac{2m}{\pi^2} \int_0^{\xi_p} d \omega
\biggl [\int_{-k_F + \sqrt{k_F^2 + 2m \omega}}^
{k_F - \sqrt{k_F^2 - 2 m \omega}} dq  \nonumber \\
&&+\int_{k_F - \sqrt{k_F^2 - 2 m \omega}}^{k_F +\sqrt{k_F^2 - 2 m \omega}} dq
+\int_{k_F +\sqrt{k_F^2 - 2 m \omega}}
^{k_F +\sqrt{k_F^2 + 2 m \omega}} dq \biggr ] \nonumber \\
&& \times q W^2(q) \frac{\Im m \chi_0(q, \omega)}
{\sqrt{4p^2q^2 - (2 m \omega+ q^2)^2}} .
\end{eqnarray}
The three regions of integration, at a given value of $\omega$ are shown in Fig.~3.
\begin{figure}\label{qintegral}
\includegraphics[width=8cm]{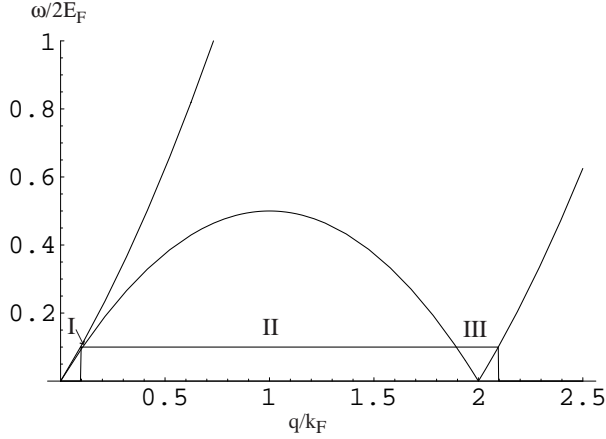}
\caption {The three regions of integration over $q$, 
at given $\omega$,  in Eq.~(\ref{threeregions}) are 
labelled as I, II and III respectively.  Only region II 
contributes to the leading order in 2D.}
\vspace{1pt}
\end{figure}
It can be shown~(\cite{Damico}) that the first and the third 
terms in the square bracket make contributions of the
order of $O(\xi_p^2)$,  
but not to the leading order of $O (\xi_p^2 \ln \xi_p)$, which arises only
from the second term. Hereafter we therefore focus only 
on the calculation of the second term.

The small $\omega$ expression for $\Im m \chi_0(q, \omega)$ in 2D is
\begin{eqnarray} \label{chi2D}
\Im m \chi_0(q, \omega)= -\frac{2 m^2 \omega}
{\pi q \sqrt{4 k_F^2 -q^2}} .
\end{eqnarray}
Therefore we obtain
\begin{eqnarray}  \label{tauD2D}
\frac{1}{\tau^{(D)}}
= \frac{4 m^3 }{\pi^3} \int_0^{\xi_p}
\omega d \omega   Q_1(\omega) ,
\end{eqnarray}
where
\begin{eqnarray}  
Q_1(\omega)  =
\int_{k_F - \sqrt{k_F^2 - 2 m \omega}}^{k_F
+\sqrt{k_F^2 - 2 m \omega}} dq 
 W^2 (q) \frac{1}{q (4 k_F^2 - q^2 )} ,
\end{eqnarray}
or, to the leading order,
\begin{eqnarray}  \label{Q1}
Q_1(\omega) = \int_{m \omega/k_F}^{2 k_F - m \omega/k_F}
dq W^2 (q) \frac{1}{q (4 k_F^2 - q^2 )} .
\end{eqnarray}
Evidently, the integration in Eq. (\ref{Q1}) has a logarithmic divergence
at both the upper and lower limits. To the leading order, $Q_1(\omega)$ can
be evaluated as
\begin{eqnarray}  \label{Q1a}
Q_1(\omega) =- \frac{1}{8 k_F^2} [ 2 W^2(0) + W^2(2k_F) ]
\ln \frac{m \omega}{k_F^2}~.
\end{eqnarray}
Substituting Eq. (\ref{Q1a}) into (\ref{tauD2D}), 
and performing the integration over $\omega$,
we finally arrive at
\begin{eqnarray} \label{2DDT=0}
\frac{1}{\tau^{(D)}}
=  \frac{\xi_p^2}{4 \pi E_F} 
\left[\bar W^2(0) + \frac{1}{2}\bar W^2(2k_F)\right]  
\ln \frac{2E_F}{\xi_p}~,
\end{eqnarray}
where we have defined the dimensionless quantity
\begin{equation}
\bar W(q) \equiv \frac{m}{\pi } W(q)~.
\end{equation}
The quantity in the square brackets of Eq.~(\ref {2DDT=0}) can 
be expressed in terms of the Wigner-Seitz radius $r_s$ as follows:
\begin{equation}
\bar W^2(0) + \frac{1}{2}\bar W^2(2k_F)
=1+\frac{1}{2}\left( \frac{r_s}{r_s +\sqrt{2}}\right)^2~.
\end{equation}
The fact that Eq. (\ref{Q1}) also has a logarithmic contribution 
from the upper limit of integration at $q \simeq 2k_F$
was missed in almost all previous analytical calculations. This is
one of the main reasons leading to errors in the numerical 
prefactor of the lifetime. The  second term in the square 
brackets of Eq.~(\ref{2DDT=0}) is absent in 
the works of Refs.~(\cite{giuliani,zheng,reizer}), 
while it is over-appreciated in the work of 
Jungwirth and MacDonald~\cite{jungwirth},  
where the factor $\frac{1}{2}$ in front of $\bar W^2(2k_F)$ is replaced by $1$.   
All these formulas are, of course, equivalent in the high-density limit ($r_s \to 0$).

Except for  the work by Reizer and Wilkins~\cite{reizer}, 
all the calculations cited above in 2D explicitly consider 
only the direct process, without taking account of the exchange process, 
which we deal with in the next subsection.

\subsection{$k_BT \ll \xi_p$: Exchange process}

In 2D, for $k_BT \ll \xi_p \ll E_F $, Eq. (\ref{tauX3}) becomes
\begin{eqnarray}
\frac{1}{\tau^{(EX)}}
=&&\frac{2m}{(2 \pi)^2} \int_0^{\xi_p} d \omega  \biggl [
\int_{-k_F + \sqrt{k_F^2 + 2m \omega}}^
{k_F - \sqrt{k_F^2 - 2 m \omega}} dq  \nonumber \\
&&+\int_{k_F - \sqrt{k_F^2 - 2 m \omega}}^{k_F +\sqrt{k_F^2 - 2 m \omega}} dq
+\int_{k_F +\sqrt{k_F^2 - 2 m \omega}}
^{k_F +\sqrt{k_F^2 + 2 m \omega}} dq \biggr ] \nonumber \\
&& \times q W(q)[W(0) + W(\sqrt{4k_F^2 - q^2})]  \nonumber \\
&&\times \frac{\Im m \chi_0(q, \omega)}
{\sqrt{4p^2q^2 -( q^2 +2 m \omega)^2}} .
\end{eqnarray}
Again only the second term in the square bracket
contributes to the leading order.
Thus,
\begin{eqnarray}  
\frac{1}{\tau^{(EX)}}  \label{tauX2D}
=- \frac{4m^3}{\pi (2 \pi)^2} \int_0^{\xi_p} d \omega \omega
Q_2 (\omega) ,
\end{eqnarray}
where
\begin{eqnarray}  \label{Q2}
Q_2 (\omega)= && \int_{k_F - \sqrt{k_F^2 - 2 m \omega}}^{k_F
+\sqrt{k_F^2 - 2 m \omega}} dq \frac{1}{q (4 k_F^2 - q^2 )}   \nonumber \\
&& \times W (q)[W(0) + W(\sqrt{4 k_F^2 - q^2 })] .
\end{eqnarray}
To the leading order,
\begin{eqnarray}  \label{Q2a}
Q_2 (\omega)= - \frac{1}{4 k_F^2} W(0)
[W(0) +2 W(2 k_F)] \ln \frac{m \omega}{k_F^2} .
\end{eqnarray}
Substituting Eq. (\ref{Q2a}) into Eq. (\ref{tauX2D}) and
performing the integration over $\omega$, we arrive at
\begin{eqnarray} \label{2DEXT=0}
\frac{\hbar}{\tau^{(EX)}}
=\frac{\xi_p^2}{16 \pi E_F} 
\bar W(0)[\bar W(0) + 2\bar W(2k_F) ]  \ln \frac{\xi_p}{2E_F}.
\end{eqnarray}
In Fig.~4, we show the ratio of $1/\tau^{(EX)}$ to $1/\tau^{(D)}$ vs $r_s$.  
The remarkable fact is that, at variance with the 3D case, 
this ratio does not vanish for $r_s \to 0$.  
The reason for this difference can be understood as follows.  
In 3D a {\it typical} scattering process near the Fermi surface, 
such as the one shown in Fig.~1(a),  involves two particles that 
are well separated (by a wave vector of the order of $2k_F$) in momentum space.  
The direct scattering amplitude for such a process is 
maximum when the momentum transfer $q$ is much smaller than $k_F$, 
and is thus typically proportional to $W(0)$.  
The exchange scattering amplitude, on the other hand, 
is of the order of $W(2k_F)$ for all values of $q$:  
hence the ratio  of $1/\tau^{(EX)}$ to $1/\tau^{(D)}$ 
goes as $\frac{W(0)W(2k_F)}{W(0)^2}$, which vanishes for $r_s \to 0$.  
The reason why this argument fails in 2D is that the 
logarithmic contribution to the inverse lifetime 
in the high density limit does not arise from typical 
scattering processes, but rather, from {\it special} ones 
in which the two colliding particles are very close 
in momentum space (see Fig.~1(b)):  hence the direct and the 
scattering amplitude are comparable, and give similar 
contributions to the inverse lifetime.
A careful analysis of the integrals involved shows that in 
the high density limit, the exchange contribution 
cancels $\frac{1}{4}$ of the direct contribution to the inverse lifetime.  
This result is at variance with that of Ref.~\cite{reizer},  
according to which the exchange contribution cancels $\frac{1}{2}$ of 
the direct one.  We find that the relation  $1/\tau^{(EX)} = -1/2\tau^{(D)}$ 
holds only in the  low density limit (see Eq. (\ref{exact2}) and Fig.~4), where 
the weak coupling theory is not reliable.

Combining direct and exchange contributions in a single formula we finally find that
\begin{eqnarray}  \label{taueeT=0}
\frac{1}{\tau_e} =  \frac{\xi_p^2}{4\pi E_F}
\left[\frac{3}{4}\bar W(0)^2 +  \frac{1}{2} \bar W(2k_F)^2 
- \frac{1}{2}\bar W(0)\bar W(2k_F) \right] \ln \frac{2 E_F}{\xi_p}~,
\end{eqnarray}
where the quantity in the square brackets is given by
\begin{equation}
\frac{3}{4}-\frac{r_s}{\sqrt{2}(r_s+\sqrt{2})^2}~. 
\end{equation}
Thus in the high density limit the total inverse 
lifetime differs by a factor $\frac{3}{4}$  from the 
result of the direct-scattering-only calculation, 
and by a factor $\frac{3}{2}$ from the result of Ref.~(\cite{reizer}). 

\begin{figure} \label{fig4}
\includegraphics[width=7cm]{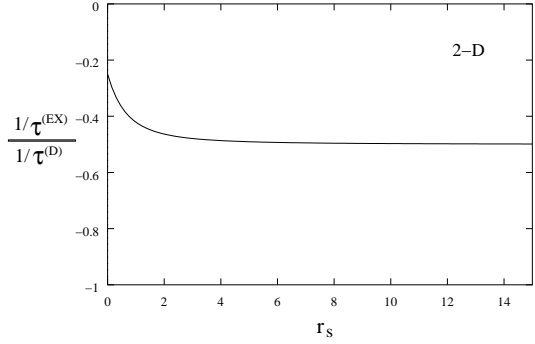}
\caption {The ratio of $1/\tau^{(EX)}$
over $1/\tau^{(D)}$ via $r_s$ in 2D.}
\vspace{1pt}
\end{figure}

\subsection{$\xi_p \ll k_BT$: Direct process}

For $\xi_p \ll k_BT \ll E_F$, Eq. (\ref{tauD3}) becomes
\begin{eqnarray}
\frac{1}{\tau^{(D)}}
= - \frac{m}{\pi^2}   
&& \biggl [ \int_{- \infty}^0 d \omega
\int_{- q_- (\omega)}^{q_+ (\omega)} d q \nonumber \\
&& +\int_0^{\mu + \xi_p} d \omega
\int_{q_- (\omega)}^{q_+ (\omega)} d q \biggl ]  \nonumber \\
&& \times \frac{1}{sh \beta \omega} \frac{q W^2(q) \Im m \chi_0(q, \omega)}
{\sqrt{4p^2 q^2 - (2 m \omega + q^2)^2}} ,
\end{eqnarray}
where $q_{\pm}(\omega)$ are the solutions of the equation
$\Omega_+ (q) = \omega$,
\begin{eqnarray}
q_{\pm}(\omega) = [p \pm \sqrt{p^2 - 2m \omega} ] .
\end{eqnarray}
Again, only the regime of $\omega \ll E_F$  contributes 
to $1/\tau_{ p}^{(D)}$ to the accuracy of the
leading order. Thus, by using Eq. (\ref{chi2D}), one has 
\begin{eqnarray}
\frac{1}{\tau^{(D)}}
= &&\frac{2m^3}{\pi^3} \int_{- \infty}^\infty
d \omega \frac{\omega}{sh \beta \omega} \biggl [
\int_{-k_F + \sqrt{k_F^2 + 2m |\omega|}}^
{k_F - \sqrt{k_F^2 - 2 m |\omega|}} dq  \nonumber \\
&&+\int_{k_F - \sqrt{k_F^2 - 2 m |\omega|}}^
{k_F +\sqrt{k_F^2 - 2 m |\omega|}} dq
+\int_{k_F +\sqrt{k_F^2 - 2 m |\omega|}}
^{k_F +\sqrt{k_F^2 + 2 m |\omega|}} dq \biggr ]    \nonumber \\
&& \times W^2(q) \frac{1}{\sqrt{4p^2 q^2 
- (2 m \omega + q^2)^2}}   \nonumber \\
&& \times \frac{1}{\sqrt{4 k_F^2 - q^2}} .
\end{eqnarray}
Once again only the second term in the bracket makes contribution
to the leading order, and we find
\begin{eqnarray}
\frac{1}{\tau^{(D)}}
= && \frac{2m^3}{\pi^3} \int_{- \infty}^\infty
d \omega \frac{\omega}{sh \beta \omega} 
Q_1(\omega) ,
\end{eqnarray}
where $Q_1 (\omega)$ is defined in Eq. (\ref{Q1}) and
evaluated in Eq. (\ref{Q1a}). Therefore
\begin{eqnarray}
\frac{1}{\tau^{(D)}}
= &&- \frac{m^3}{4\pi^3 k_F^2}
[2 W^2(0) + W^2(2 k_F) ]  \nonumber \\
&& \times \int_{- \infty}^\infty d \omega \frac{\omega}{sh \beta \omega}
\ln \frac{m \omega}{k_F^2} ,
\end{eqnarray}
which can be further evaluated leading to
\begin{eqnarray} \label{2DDT>0}
\frac{\hbar}{\tau^{(D)}}
= \frac{\left(\pi k_BT\right)^2}{8 \pi E_F}  
\left[\bar W^2(0) + \frac{1}{2}\bar W^2(2 k_F) \right] \ln \frac{2E_F}{k_BT} .
\end{eqnarray}

As in the low-temperature case, the second term in the 
square brackets of this equation  was missed in almost all
the previous theories except the one by Jungwirth and MacDonald~\cite{jungwirth}, 
which, however,  overestimates it by a factor $2$. 
Without the second term in the square bracket Eq. (\ref{2DDT>0}) would
agree with the expression obtained by Zheng and Das Sarma~\cite{zheng} and 
by Reizer and Wilkins~\cite{reizer},  
but it would be four times smaller than the result 
of Fukuyama and  Abrahams~\cite{fukuyama}, and $\frac{\pi^2}{4}$ times 
larger than the result of Giuliani and Quinn~\cite{giuliani}.

\subsection{$\xi_p \ll k_BT$: Exchange process}

In 2D, for $k_BT \gg  \xi_p$, Eq. (\ref{tauX3}) becomes
\begin{eqnarray}
\frac{1}{\tau^{(EX)}}
=&&\frac{m }{4 \pi^2} \biggl [ \int_{- \infty}^0 d \omega
\int_{- q_- (\omega)}^{q_+ (\omega)} d q   \nonumber \\
&& +\int_0^{\mu + \xi_p} d \omega
\int_{q_- (\omega)}^{q_+ (\omega)} d q \biggl ]  \nonumber \\
&& \times \frac{1}{sh\beta \omega} \frac{\Im m \chi_0(q, \omega)}
{\sqrt{4p^2 q^2 - (2 m \omega + q^2)^2}}  \nonumber \\
&& \times W(q) q [ W(0) + W (\sqrt{4 k_F^2 -q^2}) ] .
\end{eqnarray}
Proceeding as in the previous section we rewrite this as
\begin{eqnarray}
\frac{1}{\tau^{(EX)}}
= -\frac{m^3}{2\pi^3} \int_{-\infty}^\infty d \omega 
\frac{\omega}{sh \beta \omega} Q_2 (\omega) ,
\end{eqnarray}
where $Q_2 (\omega)$ is defined in Eq. (\ref{Q2}) and evaluated
in Eq. (\ref{Q2a}). Therefore
\begin{eqnarray}
\frac{1}{\tau^{(EX)}}
=&& \frac{m^3}{8 \pi^3 k_F^2} W(0)
[ W(0) +2 W(2 k_F)]  \nonumber \\
&& \times \int_{-\infty}^\infty d \omega
\frac{\omega}{sh \beta \omega} \ln \frac{m \omega}{k_F^2}~,
\end{eqnarray}
which can be, to the leading order,  further simplified to 
\begin{eqnarray} \label{2DEXT>0}
\frac{1}{\tau ^{(EX)}}
= \frac{\left(\pi k_B T\right)^2}{32 \pi E_F} \bar W(0) [\bar W(0) + 2 \bar W(2 k_F)] \ln \frac{k_B T}{2E_F}~.
\end{eqnarray}
The ratio of $1/\tau ^{(EX)}$ to $1/\tau ^{(D)}$ is therefore found
to be the same as that in the case of $k_BT \ll \xi_p$, which has been plotted in Fig.~4.

Combination of $1/\tau^{(D)}$ in Eq. (\ref{2DDT>0}) and
$1/\tau^{(EX)}$ in Eq. (\ref{2DEXT>0}) thus yields
\begin{eqnarray} \label{taueeT>0}
\frac{1}{\tau_e} 
= -\frac{\left(\pi k_B T\right)^2}{32 \pi E_F} 
[3\bar W^2(0) + 2 \bar W^2(2k_F) - 2\bar W(0)\bar W(2k_F) ] 
\ln \frac{k_B T}{2E_F}~.
\end{eqnarray}
Notice the difference between the above result and the one obtained in Ref.~\cite{reizer}.

\section{Comparison with Experimental results in 2D}

Consider  two identical 2D electron liquids in closely 
spaced quantum wells between which a small potential difference $V$ is 
maintained.  We expect a small tunneling current between the layers.  
However, as Fig.~5 shows, no tunneling is possible  
in the absence of impurities and electron interactions.  
This is because under those unrealistic assumptions both the energy 
and the momentum of the electron must be conserved during tunneling, 
and there are simply no states satisfying these conditions.    

\begin{figure}\label{tunneling}
\includegraphics[width=8cm]{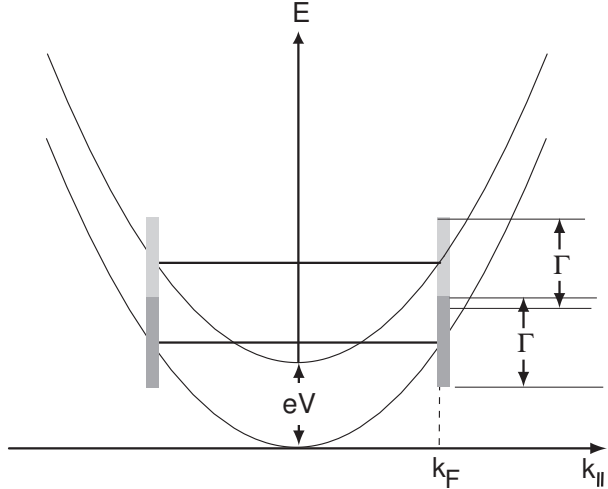}
\caption {Momentum- and energy-conserving tunneling between 
two identical free-electron bands separated by a potential difference $eV$ is 
possible only if the spectral width of the single particle states 
in each band (indicated by the shaded regions) is at least as large as  $eV$.}
\vspace{1pt}
\end{figure}

The situation changes profoundly if electron-electron 
interactions are allowed.  Now momentum is still conserved 
(if impurity scattering and surface roughness are negligible) but 
the energy of the electron quasiparticle is no longer a well 
defined quantity, due to the possibility of inelastic 
scattering processes involving other electrons in each quantum well. As a result, tunneling 
becomes possible in a region of  voltages $-\Gamma<V<\Gamma$ where $\Gamma$ 
is the width at half-maximum of the plane-wave spectral 
function in a well~\cite{murphy}, i.e.,
\begin{eqnarray}
A(E, \xi_p) = \frac{1}{2 \pi} \frac{\Gamma(\xi_p, T)}{(E-\xi_p)^2 
+ [\Gamma(\xi_p, T)/2]^2}~.  
\end{eqnarray}
From the well-known relation $A(E, \xi_p) = -\frac{1}{\pi} \Im m  G_{ret}(E, \xi_p)$, 
where $G_{ret}(E, \xi_p)$ is the retarded Green's function,  
one can show that the spectral width $\Gamma$ is just the 
inverse of the lifetime of a plane wave state, 
which is the sum of the lifetimes of electron and hole 
quasiparticles in the following manner:(\cite{jungwirth,mahan}
\begin{eqnarray}
\Gamma(\xi_p, T)  = 
\frac{1}{\tau_e(\xi_p, T)}+\frac{1}{\tau_h(\xi_p, T)} .
\end{eqnarray}
The principle of detailed balance demands
\begin{equation}
\frac{n(\xi_p,T)}{\tau_e}=\frac{1-n(\xi_p,T)}{\tau_h}~,
\end{equation}
where $n(\xi_p,T)$ is the thermal occupation number at temperature $T$.
If we assume $\xi_p\ll k_BT$ and approximate  $\Gamma(\xi_p, T)$
by $\Gamma(\xi_p=0, T)$, we see from the above equations that 
the half-width at half-maximum of
the tunneling conductance peak is expressed in terms of the 
electron quasiparticle lifetime as follows:  
\begin{eqnarray}
\Gamma= \frac{2}{\tau_e(0, T)}~.
\end{eqnarray}

We can now attempt a comparison between the experimental 
values of $\Gamma$ from Ref.~(\cite{murphy}) and 
the theoretical values of $\frac{2}{\tau_e(0, T)}$.  
This is shown in Fig.~6.  It must be kept in mind that, 
in order to perform a meaningful comparison, one must first 
subtract from the experimental data a  (presumably) 
temperature-independent constant due to residual disorder.  
The value of this constant is determined by the condition 
that $\Gamma$ tend to zero for $T \to 0$.  Even after 
this subtraction we see that the theoretical curve lies well below the experimental data.  
Furthermore, the shortcomings in Refs.~\cite{jungwirth,zheng} as revealed in this
paper imply that the  ``excellent agreement" with experiment claimed 
in those papers is  overly optimistic, as 
pointed out earlier by Reizer and Wilkins~\cite{reizer}. 

\begin{figure} \label{fig5}
\includegraphics[width=7cm]{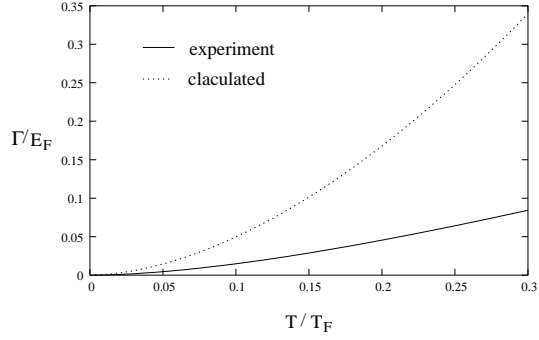}
\caption {Electron relaxation rate $\Gamma$ in 2D. Experimental data are from
Ref. \cite{murphy}, calculated ones are from Eq. (\ref{taueeT>0}). Here $T_F=E_F/k_B$.}
\vspace{1pt}
\end{figure}

We note that all derivations presented in this paper
are to the accuracy of the leading logarithmic term. Calculations including
higher order terms might bring in a better agreement with the experimental data.
However, it seems too optimistic to believe that the
huge difference (roughly a factor $4$) between theory and experiment is totally due to such higher order contributions.
The size of the discrepancy suggests that there might be other factors playing a role, like the finite
width of the quasi-two-dimensional system,  electron-impurity scattering, electron-phonon scattering, and surface roughness.  While the inclusion of these effects may help to produce better agreement with experiments, it remains a great challenge for experimentalists to device the conditions that will eventually allow them to probe the truly intrinsic behavior of the electron liquid.  

%
 \section{Acknowledgements}
We gratefully acknowledge support by NSF grants DMR-0074959 and DMR-0313681. 
We especially thank Gabriele Giuliani for many valuable discussions.

\section{Appendix}

In this appendix, we give the details of the derivation of Eq. (\ref{exaverage}).
To this end, we denote the left hand side of Eq. (\ref{exaverage}) as $A_3$ and
$A_2$ for 3D and 2D cases, respectively. In 3D, $A_3$ can be rewritten as
\begin{eqnarray}
&&A_3=e^2 \int_0^\pi d \theta' \sin \theta' \int_0^{2 \pi}d\phi'
\delta\left(\omega + \frac{pq \cos \theta'}{m} +\frac{q^2}{2m}\right)  \nonumber \\
&& \frac{1}{p^2 +k^2 +k_s^2 - 2pk [\cos \theta \cos \theta' + \sin \theta \sin \theta'
\cos(\phi - \phi')]} \nonumber \\
\end{eqnarray}
where $(\theta, \phi)$ and $(\theta', \phi')$ are the spherical angles
of ${\bf k}$ and ${\bf p}$, respectively, relative to ${\bf q}$. Carrying
through the $\phi'$ integration, one obtains
\begin{eqnarray}
&&A_3= 2 \pi e^2 \int_{-1}^1 dx \delta\left(\omega+\frac{pqx}{m} + \frac{q^2}{2m}\right)  \nonumber \\
 &&\frac{1}{\sqrt{(p^2 +k^2 + k_s^2+2pk \cos \theta x)^2-4(pk\sin \theta)^2 (1-x^2)}}  \nonumber \\
\end{eqnarray}
The integral in the above equation is trivial due to the $\delta$ function, and
it leads to
\begin{eqnarray}
A_3=  \frac{2 \pi me^2 \theta(\Omega_+ (q) - \omega)
\theta(\omega - \Omega_- (q))}{ p q \sqrt{[p^2 +k^2 +k_s^2 
- {\bf k} \cdot {\bf q} ]^2 -[k^2 - ({\bf k} \cdot {\bf \hat q})^2 ]
[4p^2 - q^2] }}\nonumber \\
\end{eqnarray}
where we have used the fact that $2m \omega\ll k_s^2$ 
Putting in this expression the approximate equalities $k \sim p \sim k_F$ and  ${\bf k} \cdot {\bf q}\sim - \frac{q^2}{2}$ (which follows from the condition $\xi_{\bf k} - \xi_{{\bf q} +{\bf k}} 
=\omega \simeq 0$ due to the first $\delta$-function in  Eq. (\ref{tauX2})) one easily arrives at Eq.~(\ref{exaverage}) in the 3D case.                                          

In 2D, $A_2$ can be explicitly written as
\begin{eqnarray}
A_2= e^2 \int_0^{2 \pi} d \phi' 
\delta(\omega +pq \cos \phi'/m + q^2/2m) \nonumber \\
\frac{1}{\sqrt{p^2 + k^2 - 2pk \cos(\phi-\phi')} +k_s}
\end{eqnarray}
or,
\begin{eqnarray}
A_2= e^2 \int_0^\pi d\phi'
\delta(\omega +pq/m \cos \phi' + q^2/2m) \nonumber \\
\biggl [\frac{1}{\sqrt{p^2 + k^2 - 2pk \cos(\phi-\phi')} +k_s} \nonumber \\
+\frac{1}{\sqrt{p^2 + k^2 - 2pk \cos(\phi+\phi')} +k_s} \biggr]
\end{eqnarray}
Carrying out the integration over $\phi'$ yields,
\begin{eqnarray}
&&A_2=\frac{me^2 \theta(\Omega_+ (q) - \omega)
\theta(\omega - \Omega_- (q))}{\sqrt{p^2q^2 -(m\omega +q^2/2)^2}} \nonumber \\
&&\left ( \frac{1}{\sqrt{p^2 +k^2 + {\bf k} \cdot {\bf q}
- \sqrt{[k^2 -({\bf k} \cdot {\bf \hat q})^2][4p^2 - q^2]}} +k_s} \right. \nonumber \\
&&\left.+ \frac{1}{\sqrt{p^2 +k^2 + {\bf k} \cdot {\bf q}
+ \sqrt{[k^2 -({\bf k} \cdot { \bf \hat q})^2][4p^2 - q^2]}} 
+k_s}\right)  \nonumber \\
\end{eqnarray} 
Substituting, as in the 3D case, the approximate equalities, 
$k \sim p \sim k_F$ and  ${\bf k} \cdot {\bf q}\sim - \frac{q^2}{2}$ one finally arrives at  Eq.~(\ref{exaverage}) in 2D.

\end{document}